\documentclass[a4paper]{jpconf}
\usepackage{graphicx}
\usepackage{amsmath}
\usepackage{amsfonts}
\usepackage{amssymb}
\usepackage{bm}
\usepackage[hmargin=2.5cm,vmargin=2.8cm]{geometry}
\usepackage{wrapfig}

\begin{document}
\title{Study of the localization-delocalization transition for phonons via transfer matrix method techniques}

\author{S D Pinski and R A Roemer}

\address{Department of Physics and Centre for Scientific Computing, University of Warwick, Coventry, CV4 7AL, UK}

\ead{s.d.pinski@warwick.ac.uk}

\begin{abstract}
We use a transfer-matrix method to study the localization properties of vibrations in a `mass and spring' model with simple cubic lattice structure. Disorder is applied as a box-distribution to the force-constants $k$ of the springs. We obtain the reduced localization lengths $\Lambda_M$ from calculated Lyapunov exponents for different system widths to roughly locate the squared critical transition frequency $\omega_{\text{c}}^2$. The data is finite-size scaled to acquire the squared critical transition frequency of $\omega_{\text{c}}^2 = 12.54 \pm 0.03$ and a critical exponent of $\nu = 1.55 \pm 0.002$.
\end{abstract}

The disorder-induced metal-insulator transition (MIT) and the concept of Anderson localization \cite{And58} for electrons has been studied extensively for many years. During this time highly accurate results for the critical disorder and critical exponent have been obtained through numerous numerical methods, such as multifractal analysis \cite{GruS95, MilRS97}, energy level statistics \cite{ZhaK97,MilRS00} and the transfer matrix method (TMM) \cite{KraM93,PicS81a}. Although phonon localization has been a topic of research for a similar duration, where the first study of a disordered phonon is credited to Lifshitz et al.\ in 1954 \cite{LifK54}, comparatively, less research has been conducted in the field. Analysis of the dynamical matrix through participation ratios \cite{CanV85,LudSTE01}, level-spacing statistics \cite{SchDG98,ShiNN07} and multifractal analysis \cite{LudTE03} requires highly accurate numerical results on large system sizes which is not trivial and remains challenging even with todays computing resources.

There are many features of interest in phononic research, such as the origins of the boson peak \cite{SchDG98,KanRB01}, the name given to the peak in the vibrational density of states (VDOS) $g(\omega)$ that is an excess contribution compared to the usual Debye behaviour $g(\omega) \propto \omega^2$. The nature of the modes within the peak are of importance as they are strongly related to the mechanisms of thermal transport in the $\approx 10 K$ temperature range \cite{SchD99}. Schirmacher et al.\ \cite{SchDG98,SchD99} successfully demonstrated that the boson peak consists of extended phonon modes. The locations of the boson peak, Ioffe-Regel transition between weak and strong elastic-scattering regimes and the localization-delocalization transition (LDT) are also of great interest in particular how these are related with each other. Scorpigno et al.\ \cite{ScoSAA06} state that the link between the locations of the boson-peak and Ioffe-Regel transition is still debatable, whereas through level statistics Kantelhardt et al.\ \cite{KanRB01} estimate the critical phonon frequency to be $\omega_{\text{c}} \approx 3 \omega_{\text{peak}}$. Taraskin et al.\ \cite{TarLNE02} declare that the links between vibrational localization, the boson peak, the Ioffe-Regel crossover and the zero-energy spectral singularity are yet to be fully established.

\begin{figure*}[ht]
\begin{center}
\begin{minipage}[t]{0.01\textwidth}
\vspace{190pt}
(a)
\end{minipage}
\hspace*{-20pt}
\begin{minipage}[t]{0.245\textwidth}
\vspace{60pt}
\includegraphics[width=0.98\textwidth]{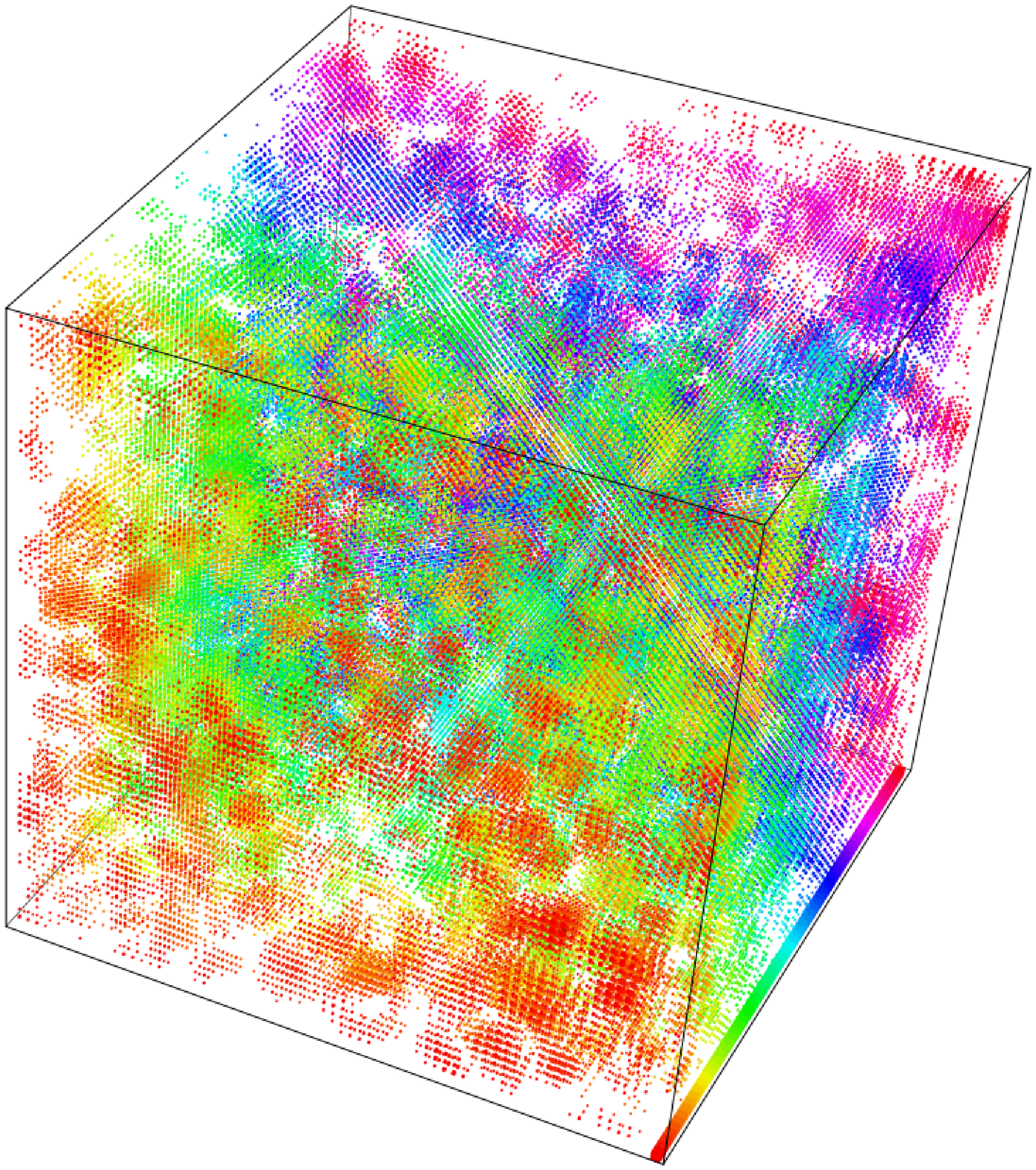}
\end{minipage}
\begin{minipage}[t]{0.01\textwidth}
\vspace{190pt}
(b)
\end{minipage}
\hspace*{3pt}
\begin{minipage}[t]{0.47\textwidth}
\vspace{0pt}
\includegraphics[width=0.98\textwidth]{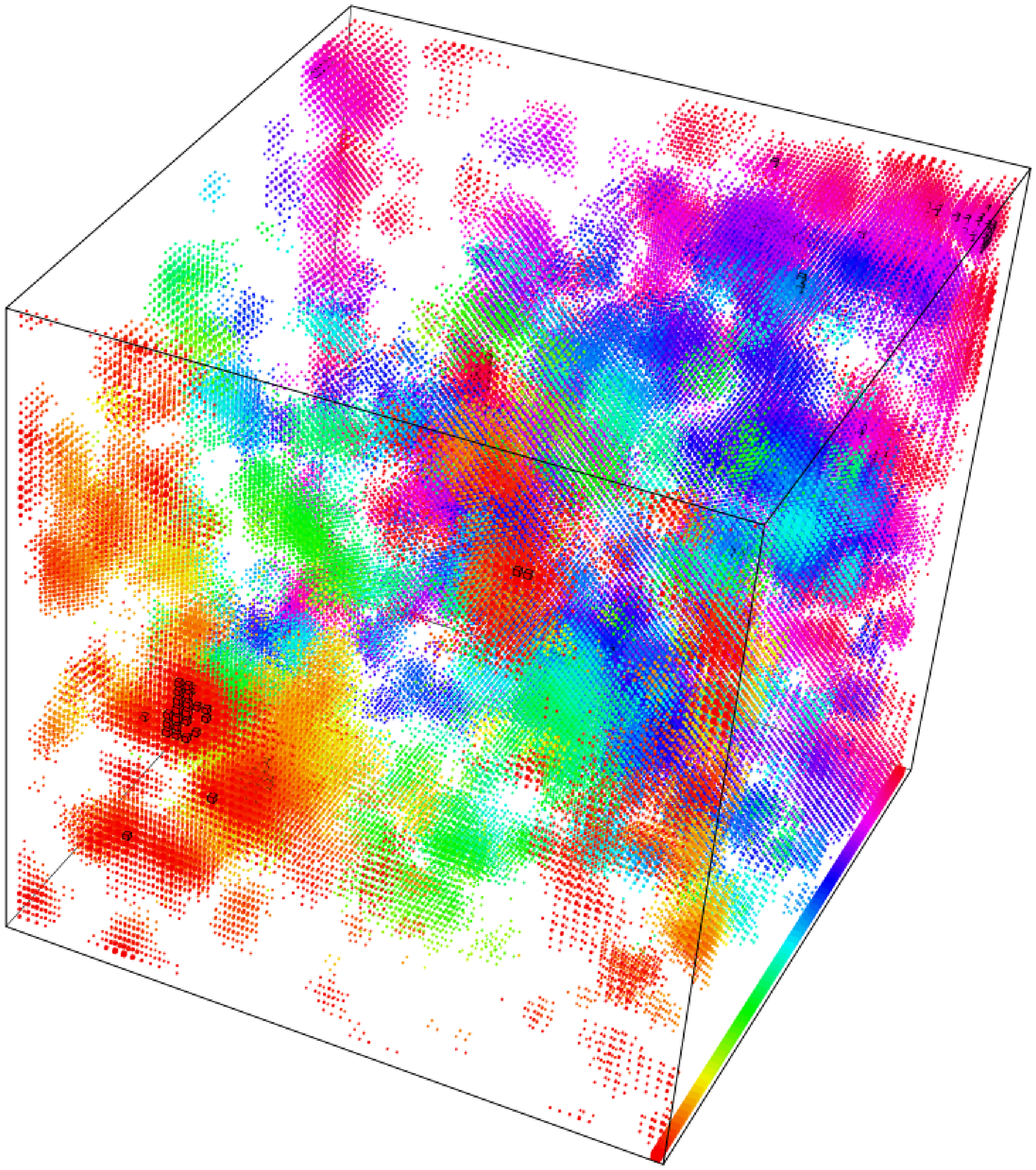}
\end{minipage}
\begin{minipage}[t]{0.01\textwidth}
\vspace{190pt}
(c)
\end{minipage}
\hspace*{-20pt}
\begin{minipage}[t]{0.245\textwidth}
\vspace{60pt}
\includegraphics[width=0.98\textwidth]{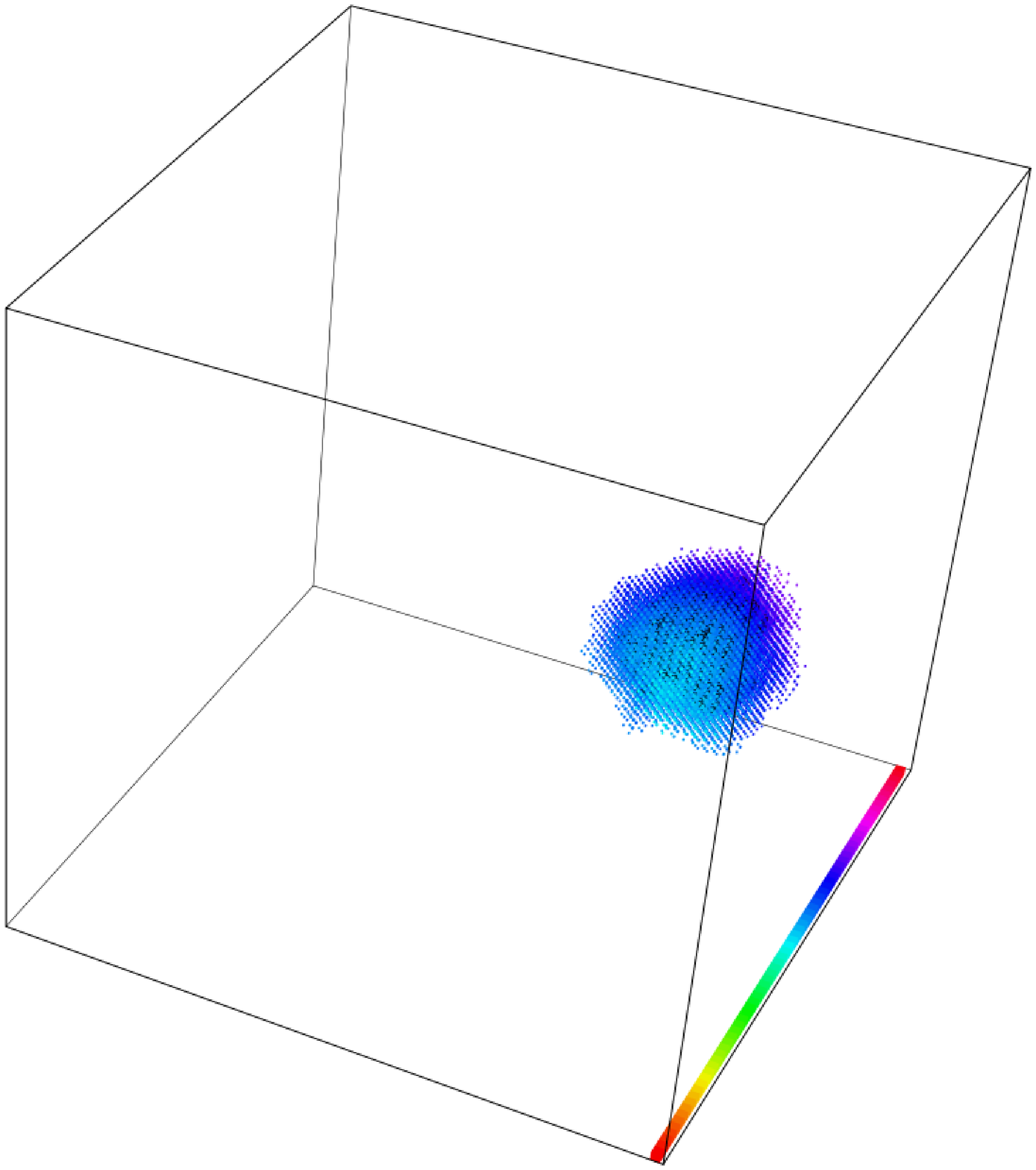}
\end{minipage}
\end{center}
\vspace*{-10pt}
\caption{Amplitudes of vibration in a box of length $N=70$ with periodic boundary conditions and evenly distributed force constant disorder with mean $\langle k \rangle =1$ and width $\Delta k = 1$.  (a) Extended state $\omega^2 = 12$, (b) critical state near the LDT transition $\omega^2 = 12.5$ and (c) localized state $\omega^2 = 13.03$. Colours are an indication of the depth within the box that a particular amplitude is situated, the scale for which is included in the lower right hand side of the boxes. Sizes of internal boxes are determined from vibrational amplitudes at lattice site. Amplitudes less than the mean are not displayed.\label{fig-states}}
\end{figure*}

Here, we take a simple cubic lattice structure made up of point masses connected by springs. We assume that the masses are displaced by only small amplitudes and apply the harmonic approximation. This gives rise to the classical model of vibration, where the displacements of the masses from their equilibrium positions are governed by the displacements of their nearest neighbours. Due to the symmetry of the lattice we can de-couple the spatial dimensions of the system into three identical problems and massively reduce the time of calculations. This is achieved by assuming that all springs have identical central and non-central force constants reducing the $3\times 3$ force constant matrices to scalars. Therefore, solving for just one spatial dimension achieves all three components of the vibrational amplitude for all lattice sites.

We apply the above approximations and hence have to solve the standard equation of motion for a single lattice site with a plane wave solution to obtain
\begin{eqnarray}
  -\omega^2 m_{l,m,n} u_{l,m,n} &=& k_{l,m,n+1} (u_{l,m,n+1} - u_{l,m,n}) + k_{l,m,n-1} (u_{l,m,n-1} -u_{l,m,n}) +\nonumber\\
&& k_{l,m+1,n} (u_{l,m+1,n} -u_{l,m,n}) + k_{l,m-1,n} (u_{l,m-1,n} -u_{l,m,n}) +\nonumber\\
&& k_{l+1,m,n} (u_{l+1,m,n} -u_{l,m,n}) + k_{l-1,m,n} (u_{l-1,m,n} -u_{l,m,n}), \label{equ-dyna}
\end{eqnarray}
where $\omega$ is the frequency of vibration, $m_{l,m,n}$ and $u_{l,m,n}$ are the mass and amplitude of vibration of the lattice site denoted by indices $l,m,n$. $k$ is the force constant of the spring connecting two adjacent lattice sites, where the indices of the current lattice site $l,m,n$ are omitted to simplify notation. The system of equations for all lattice sites is converted to matrix form by substitution into the generalised equation of motion

\begin{equation}
 -\omega^2 \mathbf{u=[m]^{-1}[k]u}
\end{equation}

The above matrices $\mathbf{[m]}$ and $\mathbf{[k]}$ contain all the masses and force constants of the eigensystem and $\mathbf{u}$ is a vector of all vibrational amplitudes of the lattice sites. $\mathbf{[m]^{-1}[k]}$ is the dynamical matrix and is diagonalized to obtain the normal modes of vibration $\mathbf{u}$ with eigenvalues $-\omega^2$. Although the vibrational amplitudes $\mathbf{u}$ are for only a single spatial dimension, they are also valid as the total magnitude of the amplitude. Recombining the individual amplitudes to obtain the total requires multiplication of single components by $\sqrt{3}$, renormalization of the eigenvector removes these factors.

Several methods can be used to analyse both the eigenvectors and eigenvalues obtained from diagonalization of the dynamical matrix to distinguish between extended and localized states. Just looking at the amplitude eigenvectors is a viable method of studying the localization lengths of the phonons, although it is not trivial due to the localization lengths diverging as the critical frequency is approached from the localized regime. Therefore in finite sized systems the localized states can appear extended as the localization length can be greater than the length of the system. We diagonalize the dynamical matrix for a box of size $70^3$ with the inclusion of box distribution disorder with width $\Delta k=1$ and mean $\langle k \rangle =1$ applied to the force constants of the springs. In figure \ref{fig-states}(a)-(c) we visualise three amplitude eigenvectors for squared frequencies of $\omega^2 = 12$, $\omega^2 = 12.5$ and $\omega^2 = 13.03$ respectively. Figure \ref{fig-states}(a) is that of an extended state as the amplitudes exhibit short range periodicity within the box and figure \ref{fig-states}(c) is that of a localized state where the amplitudes of vibration are confined to a small section of the box. Figure \ref{fig-states}(b) exhibits properties of both extended and localized states and is close to the transition.

\begin{figure}
\includegraphics[width=0.48\textwidth]{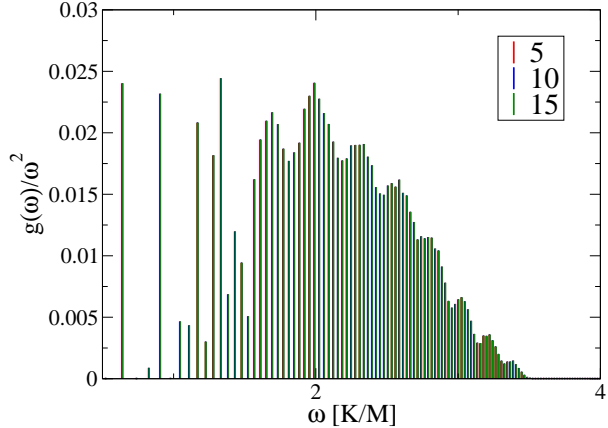}\hspace{2pc} 
\begin{minipage}[b]{14pc}\caption{Reduced VDOS for evenly distributed spring disorder with mean $\langle k \rangle =1$ and width $\Delta k=1$. 150 bins are used and the data is normalized by total number of states.\label{fig-K1.0VDOS}}
\end{minipage}
\end{figure}

We calculate the reduced VDOS $g(\omega)/\omega^2$ for cubes with size $5^3$, $10^3$ and $15^3$ with 1360, 170 and 50 random seeds respectively, resulting in roughly 170,000 states for each system size. The VDOS plot in figure \ref{fig-K1.0VDOS} demonstrates the latter stages of the broadening of the Van Hove singularities due to disorder and the emergence of the characteristic low-frequency boson peak at $\omega_{\text{peak}} \approx 2$. The frequencies $\omega^2= 12$, $12.5$ and $13.03$ of the states plotted in figure \ref{fig-states} are located close to the upper band edge of the VDOS, therefore we can safely assume that the LDT will be in this region.

We introduce the TMM which calculates the localization length $\lambda$ of a quasi-one dimensional bar. For simplicity the bar geometry has an identical height and width $M$ and in all cases the length of the bar $L \gg M$. $L$ is not preset as the transfer matrix calculations continue along the length of $L$ until the desired accuracy of the Lyapunov exponent is met, this convergence criterion is chosen before starting the calculation.

To calculate the Lyapunov exponents, the system of equations (\ref{equ-dyna}) is converted to a form suitable for the TMM. This requires the equations to be re-arranged so that the amplitudes of a particular layer of lattice sites can be calculated solely from the amplitudes of the previous layer and the force constants of the surrounding springs. Computationally this is far less memory consuming than storing the whole system of masses and springs, as the TMM calculation only requires parameters for 2 layers of the system to be stored in memory at any one time. Equation (\ref{equ-dyna}) for a single lattice site is rearranged into a form where the amplitude of vibration of the nearest neighbour in layer $l+1$ is calculated solely from parameters of sites in layers $l$ and $l-1$.

\begin{figure}
\centering
(a) \hspace*{-1cm} \includegraphics[width=0.5\columnwidth]{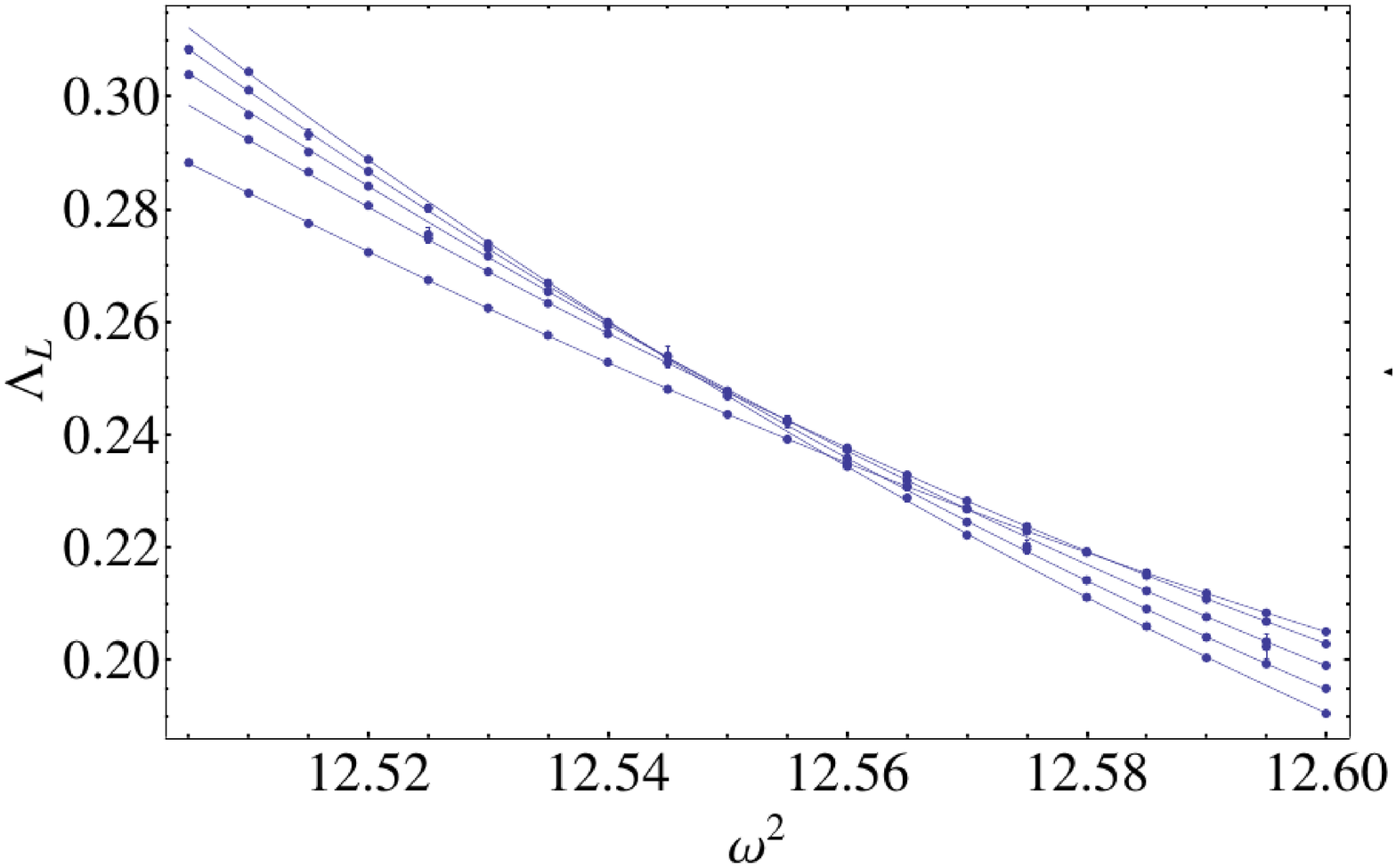} 
(b) \hspace*{-1cm} \includegraphics[width=0.5\columnwidth]{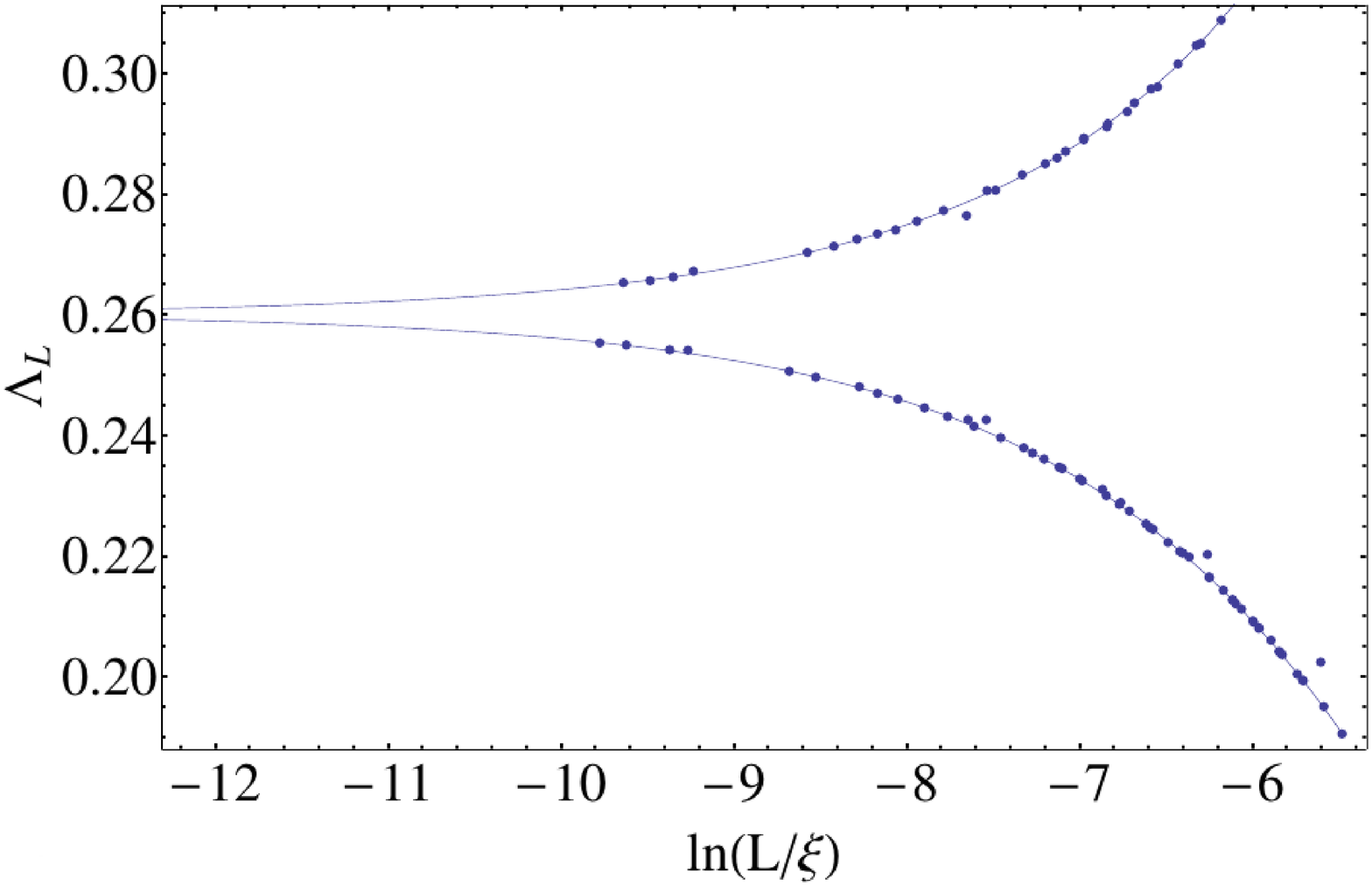} 
\caption{Spring disorder $\Delta k=1.0$: (a) reduced localization lengths for a range of frequencies and system sizes (12-20 in steps of 2). (b) Scaling function (solid line) and scaled data points for the expansion $n_{\text{r}}= 4$, $m_{\text{r}}= 1$, $n_{\text{i}}= 2$, $m_{\text{i}}= 0$.\label{fig-K1.0}}
\end{figure}

\begin{equation}
 u_{l+1,m,n} = -\frac{1}{k_{l+1,m,n}} \left[ (\omega^2 m_{l,m,n} + k_{\text{all}})u_{l,m,n} -H_{\text{l}} \right] - \frac{k_{l-1,m,n}}{k_{l+1,m,n}} u_{l-1,m,n} \label{equ-TMM_singles} 
\end{equation}

We introduce $H_{\text{l}} = k_{l,m,n+1} u_{l,m,n+1} + k_{l,m,n-1} u_{l,m,n-1} + k_{l,m+1,n} u_{l,m+1,n} + k_{l,m-1,n} u_{l,m-1,n}$ as a collection of the in-plane contributions to the final amplitude and $k_{\text{all}}$ is the sum of all surrounding force constants for site $l,m,n$. We define $U_l$, $U_{l+1}$ and $U_{l-1}$ as vectors containing the amplitudes of the constituent sites of layers $l$, $l+1$ and $l-1$ respectively. Therefore, $U_l = \left(u_{l,1,1}, u_{l,1,2}, u_{l,2,1}, \dots, u_{l,M,M} \right)$, where $M$ is the width and height of the layer. Equations (\ref{equ-TMM_singles}) is now expressed in the following matrix form, where $\mathbf{H}_{\text{l}}$ is a matrix containing all in-layer contributions, $\mathbf{0}$ and $\mathbf{1}$ are the zero and unit matrices, $\omega^2$ is single valued and all other parameters are vectors.

\begin{equation}
 \left[ \begin{array}{c}
   U_{l+1} \\ U_l
 \end{array} \right] =
\underbrace{
 \left[ \begin{array}{cc}
   -\frac{ \left[ \left(\omega^2 m_l + k_{all}\right) \mathbf{1} -\mathbf{H}_l \right] }{k_{l+1}} & -\frac{k_{l-1}}{k_{l+1}} \mathbf{1} \\ \mathbf{1} & \mathbf{0}
 \end{array} \right]}_{\mathbb{T}_n}
 \left[ \begin{array}{c}
   U_l \\ U_{l-1}
 \end{array} \right] \label{equ-1DTMM}
\end{equation}

The transfer matrix used to progress from one slice to the next is given as $\mathbb{T}_n$ and repeated multiplication of this gives: $\displaystyle \mathbb{T}_N = \prod^N_{n=1} \mathbb{T}_n$ which according to Oseledec \cite{Ose68} guarantees existence of the matrix $\displaystyle \Gamma \equiv \lim_{N\to\infty} \left(\mathbb{T}_N \mathbb{T}_N^\dagger \right)^{\frac{1}{2}N}$ with eigenvalues $e^{\gamma}$. Where $\gamma$'s are Lyapunov exponents and an estimation of localization length is given by the inverse of the minimum exponent $\lambda(M,\omega^2) = 1/\gamma_{\text{min}}$. The reduced (dimensionless) localization length may then be calculated as $\Lambda_M(\omega^2) = \lambda(M,\omega^2)/M$. We numerically calculate the reduced localization lengths for a range of squared frequencies and even system widths $M$ between 12 and 20 to a convergence criterion of $0.1$ percent of the variance. The reduced localization lengths $\Lambda_M$ as function of the squared frequency $\omega^2$ are plotted in figure \ref{fig-K1.0}(a).

The reduced localization lengths from the TMM are finite size scaled by applying the procedure outlined in reference \cite{SleO99a}. The finite size scaling (FSS) procedure obtains the correlation length $\xi$ for an infinite system from the reduced localization length $\Lambda_M (\omega^2)$ of finite sized systems by using the one-parameter scaling law $\Lambda_M = f(M / \xi)$ \cite{Tho74}. We assume that the LDT is characterised by a divergent correlation length, so that at a fixed disorder $\Delta k$, $\xi(\omega^2) \propto \mid \omega^2 - \omega_{\text{c}}^2 \mid ^{-\nu}$, where $\nu$ is the critical exponent and $\omega_{\text{c}}^2$ is the critical frequency squared. Highly accurate numerical studies of the Anderson model for electron localization have found the critical exponent $\nu \equiv 1.5 \pm 0.1$ \cite{MacK81, Mac94, SleO99a} for box distribution disorder. Whereas in the phononic case Monthus and Garel \cite{MonG10} have demonstrated that their participation ratio data for high disorder at an LDT collapsed fairly well onto their scaling function using a critical exponent $\nu = 1.57$ taken from reference \cite{EveM08}. Akita and Ohtsuki \cite{AkiO98} find a critical exponent of $\nu \approx 1.2 \pm 0.2$ by finite size scaling TMM data at an LDT for box distribution spring disorder $\Delta k = 1.8$ with a convergence criterion of 2 percent. Table \ref{tab-K1.0} contains the critical parameters obtained from FSS for particular orders of Taylor expansions of the relevant and irrelevant scaling variables $n_{\text{r}}$, $n_{\text{i}}$, $m_{\text{r}}$ and $m_{\text{i}}$ listed in columns 1-4. Figure \ref{fig-K1.0}(b) is the scaling function obtained from the highest order stable fit given in table \ref{tab-K1.0} ($n_{\text{r}}=4$, $n_{\text{i}}=1$, $m_{\text{r}}=2$, $m_{\text{i}}$=0). The mean critical exponent $\langle \nu \rangle = 1.55 \pm 0.002$ which is remarkably close to the MIT critical exponent.

\begin{table}
\caption{\label{tab-K1.0}Table of critical parameters obtained from finite size scaling for $\Delta k=1$ on even numbered system widths between $12$ and $20$ and a range of squared frequencies $\omega_{\text{c}}^2=12.5 - 12.6$. NDF is the number of degrees of freedom. We note that preliminary errors for $\nu$ are too small as they have not been fully checked with respect to stability and robustness.}
\centering
\begin{tabular}{lllllllll}
\br
$n_{\text{r}}$ & $n_{\text{i}}$ & $m_{\text{r}}$ & $m_{\text{i}}$ & $\omega_{\text{c}}^2$ & $\nu$ & $\chi^2$ & NDF & $\Gamma_q$ \\
\mr
 3 & 1 & 1 & 0 & $12.538 \pm 0.05$ & $1.5023 \pm 0.002$ & 51.12 & 80 & 0.995 \\
 3 & 1 & 2 & 0 & $12.540 \pm 0.07$ & $1.5507 \pm 0.002$ & 48.13 & 79 & 0.997 \\
 3 & 1 & 3 & 0 & $12.540 \pm 0.08$ & $1.5782 \pm 0.002$ & 46.71 & 78 & 0.998 \\
 4 & 1 & 2 & 0 & $12.540 \pm 0.07$ & $1.5573 \pm 0.002$ & 47.62 & 77 & 0.997 \\
\br
\end{tabular}
\end{table}

From FSS we achieve a weighted mean critical squared transition frequency of $\omega_{\text{c}}^2 = 12.54 \pm 0.03$ which as we can see from figure \ref{fig-K1.0VDOS} occurs in the tail of the VDOS and close to the example state of squared frequency $\omega_{\text{c}}^2 = 12.5$ in Figure \ref{fig-states}(b) that exhibits properties of both extended and localized states. It is generally assumed that the LDT shares the same exponent as the MIT, where the current accepted value is $\nu \approx 1.5 \pm 0.1$. Although Akita and Ohtsuki \cite{AkiO98} have previously obtained a much lower value $\nu \approx 1.2 \pm 0.2$, thought to be a consequence of the transition occurring close to the upper band edge \cite{KawKO99},  we have achieved a striking resemblance between critical exponents between the MIT and LDT. Therefore we assume that the phononic model will be of the same orthogonal universality class as the Anderson model.
\newline \newline
We gratefully acknowledge EPSRC for financial support (EP-F040784-1).

\vspace*{-2pt}
\section*{References}

\providecommand{\newblock}{}


\begin{thebibliography}{10}
\expandafter\ifx\csname url\endcsname\relax
  \def\url#1{{\tt #1}}\fi
\expandafter\ifx\csname urlprefix\endcsname\relax\def\urlprefix{URL }\fi
\providecommand{\eprint}[2][]{\url{#2}}

\bibitem{And58}
Anderson P~W 1958 {\em Phys. Rev.\/} {\bf 109} 1492--1505

\bibitem{GruS95}
Grussbach H and Schreiber M 1995 {\em Phys. Rev. B\/} {\bf 51} 663--666

\bibitem{MilRS97}
Milde F, {R\"{o}mer} R~A and Schreiber M 1997 {\em Phys. Rev. B\/} {\bf 55}
  9463--9469

\bibitem{ZhaK97}
Zharekeshev I~K and Kramer B 1997 {\em Phys. Rev. Lett.\/} {\bf 79} 717--720
  {ArXiv}: cond-mat/9706255

\bibitem{MilRS00}
Milde F, {R\"{o}mer} R~A and Schreiber M 2000 {\em Phys. Rev. B\/} {\bf 61}
  6028--6035 {ArXiv}: cond-mat/9909210

\bibitem{KraM93}
Kramer B and MacKinnon A 1993 {\em Rep. Prog. Phys.\/} {\bf 56} 1469--1564

\bibitem{PicS81a}
Pichard J~L and Sarma G 1981 {\em J. Phys. C\/} {\bf 14} L127--L132

\bibitem{LifK54}
Lifshitz I and Kosevich A 1954 {\em Journal of Physics: USSR\/} {\bf 8}
  217--254

\bibitem{CanV85}
Canisius J and {van Hemmen} J 1985 {\em J. Phys. C\/} {\bf 18} 4873--4884

\bibitem{LudSTE01}
Ludlam J, Stadelmann T, Taraskin S and Elliott S 2001 {\em Journal of
  Non-Crystalline Solids\/} {\bf 293} 676--681

\bibitem{SchDG98}
Schirmacher W, Diezemann G and Ganter C 1998 {\em Phys. Rev. Lett.\/} {\bf 81}
  136--139

\bibitem{ShiNN07}
Shima H, Nishino S and Nakayama T 2007 {\em Journal of Physics: Conference
  Series\/} {\bf 92} 012156

\bibitem{LudTE03}
Ludlam J, Taraskin S and Elliott S 2003 {\em Phys. Rev. B\/} {\bf 67} 132203

\bibitem{KanRB01}
Kantelhardt J~W, Russ S and Bunde A 2001 {\em Phys. Rev. B\/} {\bf 63}

\bibitem{SchD99}
Schirmacher W and Diezemann G 1999 {\em Ann. Phys. (Leipzig)\/} {\bf 8}
  727--732

\bibitem{ScoSAA06}
Scopigno T, Suck J, Angelini R, Albergamo F and Ruocco G 2006 {\em Phys. Rev.
  Lett.\/} {\bf 96} 135501

\bibitem{TarLNE02}
Taraskin S, Ludlam J, Natarajan G and Elliott S 2002 {\em Philosophical
  Magazine B\/} {\bf 82} 197--208

\bibitem{Ose68}
Oseledec V~I 1968 {\em Trans. Moscow Math. Soc.\/} {\bf 19} 197--231

\bibitem{SleO99a}
Slevin K and Ohtsuki T 1999 {\em Phys. Rev. Lett.\/} {\bf 82} 382--385 {ArXiv}:
  cond-mat/9812065

\bibitem{Tho74}
Thouless D~J 1974 {\em Phys. Rep.\/} {\bf 13} 93--142

\bibitem{MacK81}
MacKinnon A and Kramer B 1981 {\em Phys. Rev. Lett.\/} {\bf 47} 1546--1549

\bibitem{Mac94}
MacKinnon A 1994 {\em J. Phys.: Condens. Matter\/} {\bf 6} 2511--2518

\bibitem{MonG10}
Monthus C and Garel T 2010 {\em Phys. Rev. B\/} {\bf 81} 224208

\bibitem{EveM08}
Evers F and Mirlin A~D 2008 {\em Rev. Mod. Phys.\/} {\bf 80} 1355--1417

\bibitem{AkiO98}
Akita Y and Ohtsuki T 1998 {\em J. Phys. Soc. Jap.\/} {\bf 67} 2954--2955

\bibitem{KawKO99}
Kawarabayashi T, Kramer B and Ohtsuki T 1999 {\em {Ann. Phys. (Leipzig)}\/}
  {\bf 8} 487--496 {ArXiv}: cond-mat/9907319

\end{thebibliography}
\end{document}